\documentstyle[12pt]{JHEP3}
\newtheorem{theorem}{Theorem}
\newtheorem{proposition}{Proposition}

\title{Hyperbolic Space Forms and Orbifold Compactification in M-Theory}

\author{Andrey A. Bytsenko \\
       Depto. de F\'{\i}sica, Univ. Estadual de Londrina, Paran\'a, Brazil\\
        E-mail: abyts@uel.br}

\author{Maria Em\'{\i}lia X. Guimar\~aes\\
       Depto. de Matem\'atica, Univ. de Bras\'{\i}lia, DF,
       Brazil\\
       E-mail: marg@unb.br}

\author{Jos\'e Abdalla Helayel-Neto\\
        Centro Brasileiro de Pesquisas Fisicas, Rio de Janeiro, Brazil\\
        E-mail: helayel@cbpf.br}

\abstract{We analyze solutions of string theory and supergravity
which involve real hyperbolic spaces. Examples of string
compactifications are given in terms of hyperbolic coset spaces of
finite volume $\Gamma\backslash {\mathbb H}^N$, where $\Gamma$ is
a discrete group of isometries of ${\mathbb H}^N$. We describe
finite flux and the tensor kernel associated with
hyperbolic spaces. The case of arithmetic
geometry of $\Gamma = SL(2, {\mathbb Z}+i{\mathbb Z})/\{\pm Id\}$,
where $Id$ is the identity matrix, is analyzed. We discuss
supersymmetry surviving for supergravity solutions
involving real hyperbolic space factors, string-supergravity
correspondence and holography principle for a class of conformal
field theories.}

\begin{document}


\section{Introduction}

In theory of supergravity and string theory the de Sitter,
anti de Sitter spaces and $N-$spheres ${\mathbb S}^N$ play
important role. These spaces as well as
$N-$dimensional real hyperbolic spaces ${\mathbb H}^N$ naturally
arise as the near-horizon of black brane geometries. Spheres and
the anti de Sitter spaces, as
supergravity solutions, have been extensively investigated (see,
for example, \cite{DNP}). As for the de Sitter spaces, not much
calculations has been done for such of solutions.
The reasons are: the de Sitter spaces break
supersymmetry, and do not describe universes with zero
cosmological constant.

The $N-$dimensional real hyperbolic space can be represent as the
symmetric space $X=G/K$, where $G=SO_1(N,1)$ and $K=SO(N)$ is a
maximal compact subgroup of $G$. Hyperbolic space forms admit
Killing spinors \cite{BF,FY,LP,Lu} but they have infinite volume,
and do not seem useful for describing internal spaces in string
compactifications. Let us regard $\Gamma$ as a discrete subgroup
of $G$ acting isometrically on $X$, and take $X_{\Gamma}$ to be
quotient space by that action: $X_{\Gamma} = \Gamma\backslash
G/K$. The question of interest is wether space
$X_{\Gamma}$ admits Killing spinors and preserves supersymmetry.
Comments on this question for some exaples of finite volume
hyperbolic spaces the reader can found in \cite{Russo}.
There is a large class of new string regular compactifications
(except possible orbifold points) involving real hyperbolic spaces
or its coset spaces with very small $\alpha '$ string corrections.
Among the possible applications of these solutions, the definition
of new ${\cal N}=0$ conformal field theories in four dimensions
via holography \cite{Russo} (see also \cite{kach,NV}), and new
models for dimensional reduction in non-compact spaces.

In this paper we discuss
solutions of the eleven-dimensional supergravity which can be
presented by means of direct product of spaces containing real
hyperbolic space forms as factors. We derive
the Laplace operator on $p-$forms, the trace formula applied to
the tensor kernel and spectral functions of compact hyperbolic
spaces. We take into account the orbifolding of the
discrete group $\Gamma$. Finally we discuss the
questions of supersymmetry surviving under the orbifolding,
and string-supergravity correspondence in
its connection to the holographic principle.

\section{Hyperbolic sectors in eleven-dimensional backgrounds}

In eleven-dimensional supergravity the graviton multiplet contains
the graviton ${\rm g}_{MN}$, the antisymmetric three-form
$A_{MNK}$ and the gravitino $\Psi_M$ ($M,N,K, ...=0,1, ...,10$).
The bosonic part of the supergravity Lagrangian has the form
\begin{eqnarray}
{\mathcal L}_{({\rm boson})} & = &
\frac{1}{2\kappa_{11}^2}\sqrt{{\rm g}} \left(R-\frac{1}{2\cdot
4!}F_{MNPQ}F^{MNPQ}\right)
\nonumber \\
& - & \frac{1}{{12}^4}\varepsilon^{M_1\ldots M_{11}} A_{M_1M_2M_3}
F_{M_4\ldots M_7}F_{M_8\ldots M_{11}} \, .
\label{action}
\end{eqnarray}
A solution to the equations of motions
\begin{eqnarray}
R_{MN} & = &\frac{1}{12}\left(F_{MPQR}{F_N}^{PQR}-\frac{1}{12}
{\rm g}_{MN}F^2\right)\ ,
\label{Eins} \\
\nabla_M F^{MNPQ} & = &-\frac{1}{1152}
\varepsilon^{NPQR_1...R_8}F_{R_1...R_4} F_{R_5...R_8}\, ,
\label{F}
\end{eqnarray}
is provided by the Freund-Rubin ansatz for the antisymmetric field
strength
\begin{equation}
F_{mnpq} =
6m_0\,\varepsilon_{mnpq}\,\,\,\,\,\mbox{for}\,\,\,\,\,
m,n,...=7,...,10\,; \,\,\,\,\,\,\,\,\,\,\,\,\,\,\,
F_{MNPQ} = 0\,, \,\,\,\,\,\,\, \mbox{otherwise}\,.
\label{FR}
\end{equation}
By substituting this ansatz into the field equations (\ref{Eins})
we get
\begin{equation}
R_{\mu\nu} =  -6m^2_0 {\rm g}_{\mu\nu}\,, \,\,\,\,\,\,\,
\mu,\nu=0,...6\,,\,\,\,\,\,\,\,\,\,\,\,\,
R_{mn} =  12m^2_0 {\rm g}_{mn}\,.
\label{sol}
\end{equation}
The requirement of unbroken supersymmetry, i.e., the vanishing of
the gravitino  transformation
\begin{equation}
\delta\Psi_M=\nabla_M \varepsilon-\frac{1}{288}
\left({\Gamma_M}^{PQRS}-8
{\delta_M}^P\Gamma^{QRS}\right)\varepsilon F_{PQRS}\,,
\end{equation}
for the ansatz (\ref{FR}), is equivalent to the existence of
$SO(1,6)$ and $ SO(4)$ Killing spinors $\theta$ and $\eta$,
respectively, which satisfy
\begin{equation}
\nabla_{\mu} \theta = \pm {1\over 2} m_0 \gamma_{\mu} \theta\, ,
\,\,\,\,\,\,\,\,\,\,
\nabla_m \eta =  \pm m_0 \gamma_m \eta\, ,
\label{sf}
\end{equation}
where $\gamma_{\mu}(\gamma_m)$ are $SO(1,6)~(SO(4))$
$\gamma$-matrices. Eq. (\ref{sol}) admits solution of the form
$X^7\times Y^4$ where $X^7$ and $Y^4$ are Einstein spaces of
negative and positive  curvature, respectively. But only  those
spaces that admit  Killing spinors obeying
Eq.~(\ref{sf}) preserve supersymmetry. The
integrability conditions of Eq. (\ref{sf}) are
$ W_{\mu\nu\rho\sigma}\gamma^{\rho\sigma}\theta = 0 $, $
W_{mnpq}\gamma^{pq}\eta = 0, $ where $W_{\mu\nu\rho\sigma},
W_{mnpq}$ are the Weyl tensors of $X^7,~Y^4$, respectively. Thus,
obvious supersymmetric examples for $Y^4$ include the round
four-sphere ${\mathbb S}^4$ and its orbifolds $\Gamma\backslash {\Bbb
S}^4$, where $\Gamma$ is an appropriate discrete group
\cite{FKPZ}. For the $X^7$ space one can take the anti de Sitter
space $AdS_7$, which preserves supersymmetry as well, and leads to
the $AdS_7\!\times\!{\mathbb S}^4$ vacuum of eleven-dimensional
supergravity. There are solutions to Eq.~(\ref{sol}) involving
hyperbolic spaces which are vacua of eleven-dimensional
supergravity, and solve Eqs. (\ref{Eins}), (\ref{F}):
\begin{itemize}
\item (i)\,\,\,\,\,\,\,\,\, $AdS_{7-N}\!\times\!
{\mathbb  H}^{N}\!\times\!{\mathbb S}^4$, \, $N \geq 2$ \item
(ii)\,\,\,\,\,\,\, $AdS_{3}\!\times\!{\mathbb H}^{2}\!\times\!
{\mathbb H}^2\!\times\!{\mathbb S}^4$ \item (iii)\,\,\,\,\,
$AdS_{2}\!\times\!{\mathbb H}^{2}\!\times\!{\mathbb H}^3\!\times\!
{\mathbb S}^4$
\end{itemize}

\section{Hyperbolic geometry in type II supergravity}

\subsection{Fluxes on $G/K$}

Note that a conformal field theory involving the upper half
three-space ${\mathbb H}^3$ can be constructed as a WZW model
based on the coset $SL(2,{\mathbb C})/SU(2)$. In particular, this
theory can be combined with the NS5-brane by summing both
conformal sigma models. In the time
direction one needs to add a linear dilaton in order to saturate
the central gauge. Such a construction leads to an NS-NS two-form
gauge field with imaginary components. But by S-duality it could
be converted into a R-R two-form with imaginary components, which
leads to a solution of type IIB* theory. Thus, as a result the
conformal model is an exact solution of string theory to all
orders in the $\alpha'- $expansion. These sigma models can also be
constructed directly by brane intersections \cite{Russo,tseytlin}.
To construct solutions with brane charges, we could start with the
NS5 brane, with flux on ${\mathbb H}^3$, which is a formal analog
of the NS five brane solution, in which the three-sphere is
replaced by a hyperbolic space. By S-duality, and redefining the
RR two-form, this
result could be converted into a solution of type IIB*
supergravity.

Near horizon solution of the discussing model is a direct product
of flat space with linear dilaton and a $SL(2,{\mathbb C})/SU(2)$
WZW model. Also the near-horizon geometry of the NS5 brane with
flux on ${\mathbb H}^3$ describes a background $AdS_3\times
{\mathbb S}^3\times {\mathbb H}^3\times S^1$, with a linear
dilaton in the time direction \cite{Russo}. By U-duality, one can
construct also different D-brane solutions with time dependence.
In order to have finite flux, the space ${\mathbb H}^3$ can be
replaced by the finite volume $\Gamma \backslash {\mathbb H}^3$
space.

\subsection{Cusp forms}
We begin with the Dirac heat kernel ${\rm
Tr}({\mathfrak D}e^{-t{\mathfrak D}})$ which could be expanding as
a series of orbital integrals associated to the conjugacy classes
$[\gamma]$ in $\Gamma$. Each orbital integral, over a necessary
semisimple orbit, can be in turn expressed in terms of the
noncommutative Fourier transform of the heat kernel, along the
tempered unitary dual of $G$, the group of isometries of the
symmetric space $G/K$. The results of \cite{Moscovici1} on the
series expansion for the case of compact locally symmetric spaces
of higher ranks has been extended to the odd dimensional
non-compact spaces with cusps in \cite{Park}. More precisely,
taking into account the fixed Iwasawa decomposition $G=KAN$,
consider a $\Gamma-$cuspidal minimal parabolic subgroup $G_P$ of
$G$ with the Langlands decomposition $G_P=BAN$, $B$ being the
centralizer of $A$ in $K$. Let us define the Dirac operator
${\mathfrak D}$, assuming a spin structure for $\Gamma \backslash
{\rm Spin}(2k+1, 1)/{\rm Spin}(2k+1))$. The spin bundle
$E_{\tau_s}$ is the locally homogeneous vector bundle defined by
the spin representation $\tau_s$ of the maximal compact group
${\rm Spin}(2k+1)$. One can decompose the space of sections of
$E_{\tau_s}$ into two subspaces, which are given by the half spin
representations $\sigma_{\pm}$ of ${\rm Spin}(2k)\subset {\rm
Spin}(2k+1)$. Let us consider a family of functions ${\mathcal
K}_t$ over $G= {\rm Spin}(2k+1, 1)$, which is given by taking the
local trace for the integral kernel $\exp(-t{\mathfrak D}^2)$ (or
${\mathfrak D}\exp(-t{\mathfrak D}^2)$). The Selberg trace formula
applied to the scalar kernel function ${\mathcal K}_t$ holds
\cite{Park}:
\begin{eqnarray}
\sum_{\sigma = \sigma_{\pm}}\sum_{\lambda_k\in \sigma_p^{\pm}}
{\hat {\mathcal K}}_t(\sigma, i\lambda_k) & - &
\frac{i}{4\pi}\int_{\mathbb R}\!\!ds {\rm
Tr}\left(S_{\Gamma}(\sigma_{\pm},-s)
\frac{d}{ds}S_{\Gamma}(\sigma_{+},s) \pi_{\Gamma}(\sigma_{+},s)
({\mathcal K}_t)\right)
\nonumber \\
& = & I_{\Gamma}({\mathcal K}_t) + H_{\Gamma}({\mathcal K}_t) +
U_{\Gamma}({\mathcal K}_t)\, \mbox{,} \label{trace}
\end{eqnarray}
where $\sigma_p := \sigma_p^{+} \bigcup \sigma_p^{-}$ gives the
point spectrum of $\mathfrak D$, $S_{\Gamma}(\sigma_{+},
i\lambda)$ is the intertwining operator and $I_{\Gamma}({\mathcal
K}_t)$,  $H_{\Gamma}({\mathcal K}_t)$, $U_{\Gamma}({\mathcal
K}_t)$ are the identity, hyperbolic and unipotent orbital
integrals. If ${\mathcal K}_t$ is given by ${\mathfrak
D}e^{-t{\mathfrak D}^2}$, then $I_{\Gamma}({\mathcal K}_t)=0$ by
the Fourier transformation of ${\mathcal K}_t$. The analysis of
the unipotent orbital integral $U_{\Gamma}({\mathcal K}_t)$ gives
the following result \cite{Barbasch,Park}: All of the unipotent
terms vanish in the Selberg trace formula applied to the odd
kernel function ${\mathcal K}_t$ given by ${\mathfrak
D}e^{-t{\mathfrak D}^2}$. It means that one can obtain spectral
invariants in the case of cusps similar to that in the case of
smooth compact odd dimensional manifolds.

\section{The arithmetic geometry of
$\Gamma = SL(2,{\mathbb Z}+i{\mathbb Z})/\{\pm Id\}$}

Let $\tau$ be an irreducible representation of $K$ on a complex
vector space $V_\tau$, and form the induced homogeneous vector
bundle $G\times_K V_\tau$. Restricting the $G$ action
to $\Gamma$ we obtain the quotient bundle $E_\tau=\Gamma\backslash
(G\times_KV_\tau)\longrightarrow X_{\Gamma}=\Gamma\backslash X$
over $X$.
The natural Riemannian structure on $X$ (therefore on $X_{\Gamma}$)
induced
by the Killing form $(\;,\;)$ of $G$ gives rise to a connection
Laplacian ${\mathfrak L}$ on $E_\tau$. If $\Omega_K$ denotes the
Casimir operator of $K-$that is $ \Omega_K=-\sum y_j^2, $ for a
basis $\{y_j\}$ of the Lie algebra ${\mathfrak k}_0$ of $K$, where
$(y_j\;,y_\ell)=-\delta_{j\ell}$, then
$\tau(\Omega_K)=\lambda_\tau{\mathbf 1}$ for a suitable scalar
$\lambda_\tau$. Moreover for the Casimir operator $\Omega$ of $G$,
with $\Omega$ operating on smooth sections $\Gamma^\infty E_\tau$
of $E_\tau$ one has
$
{\mathfrak L}=\Omega-\lambda_\tau{\mathbf 1}\,.
$
For $\lambda\geq 0$ let
$
\Gamma^\infty\left(X_{\Gamma}\;,E_\tau\right)_\lambda=
\left\{s\in\Gamma^\infty E_\tau\left|-{\mathfrak L}s=\lambda s\right.
\right\}
$
be the space of eigensections of ${\mathfrak L}$ corresponding to
$\lambda$. Here we note that if $X_{\Gamma}$ is compact we can
order the spectrum of $-{\mathfrak L}$ by taking
$0=\lambda_0<\lambda_1<\lambda_2<\cdots$;
$\lim_{j\rightarrow\infty}\lambda_j=\infty$. We shall specialize
$\tau$ to be the representation $\tau^{(p)}$ of $K=SO(N)$ on
$\Lambda^p {\mathbb C}^{N}$. It will be convenient moreover to work
with the normalized Laplacian ${\mathfrak L} = -c(N){\mathfrak L}$ where
$c(N)=2(N-1)=2(2k-1)$. ${\mathfrak L}$ has spectrum
$\left\{c(N)\lambda_j\;,m_j\right\}_{j=0}^\infty$ where the
multiplicity $m_j$ of the eigenvalue $c(N)\lambda_j$ is given by $
m_j={\rm dim}\;\Gamma^\infty\left(X_{\Gamma}\;,E_{\tau^{(p)}}
\right)_{\lambda_j}.$

Let us consider group of
local isometry associated with a simple three-dimensional complex
Lie group. The discrete group can be chosen in the form $\Gamma
\subset PSL(2, {\mathbb C})\equiv SL(2,\mathbb C)/\{\pm Id\}$,
where $Id$ is the $2\times 2$ identity matrix and is an isolated
element of $\Gamma$. The group $\Gamma$ acts discontinuously at
point $z\in\bar{\mathbb C}$, $\bar{\mathbb C}$ being the extended
complex plane. We consider a special discrete group $SL(2,{\mathbb
Z}+i{\mathbb Z})/\{\pm Id\}$, where $\mathbb Z$ is the ring of
integer numbers. The element $\gamma\in\Gamma$ will be identified
with $-\gamma$. The group $\Gamma$ has, within a conjugation, one
maximal parabolic subgroup $\Gamma_\infty$.
Let us consider an arbitrary integral operator with kernel
$k(z,z')$. Invariance of the operator is equivalent to fulfillment
of the condition
$k(\gamma z,\gamma z') = k(z,z')$ for any
$z,z'\in {\mathbb H}^3$ and $\gamma\in
PSL(2,{\mathbb C})$. So the kernel of the invariant operator is a
function of the geodesic distance between $z$ and $z'$. It is
convenient to replace such a distance with the fundamental
invariant of a pair of points $u(z,z')=|z-z'|^2/yy'$, thus
$k(z,z')=k(u(z,z'))$ . Let $\lambda_j$ be the isolated eigenvalues
of the self-adjoint extension of the Laplace operator and let us
introduce a suitable analytic function $h(r)$ and
$r^2_j=\lambda_j-1$. It can be shown that $ h(r)$ is related to
the quantity $k(u( z,\gamma z))$ by means of the Selberg
transform. Let us denote by $g(u)$ the Fourier transform of $
h(r)$, namely $g(u)= (2\pi)^{-1}\int_{\mathbb
R}dr\,h(r)\exp(-iru)$.

\begin{theorem}
Suppose $h(r)$ to
be an even analytic function in the strip $|\Im r|<1+\varepsilon $
($\varepsilon>0$), and $h(r)={\mathcal O}(1+|r|^2)^{-2}$.
For the special discrete group
$SL(2,{\mathbb Z}+i{\mathbb Z})/\{\pm Id\}$
the Selberg trace formula holds
\begin{eqnarray}
\sum_j h(r_j) & - &
\sum_{\scriptstyle\{\gamma\}_{\Gamma},\gamma\not=Id,
\atop\scriptstyle\gamma-non-parabolic}\int d\mu(z)\, k(u(z,\gamma
z))
\nonumber \\
& - & \frac{1}{4\pi}\int_{\mathbb R} dr\, h(r)
\frac{d}{ds}{\log}\,S(s)|_{\atop s=1+ir} + \frac{h(0)}{4}[S(1) -1]
-Cg(0)
\nonumber \\
& = & {\rm Vol}(\Gamma\backslash G)\!\int_0^\infty \! \frac{dr\,
r^2}{2\pi^2}\:h(r) -\frac{1}{4\pi}\!\int_{\mathbb R}\!\! dr h(r)
\psi(1+ir/2) \mbox{.} \label{Selberg}
\end{eqnarray}
\end{theorem}
\noindent
The first term in the right hand site of
Eq.~(\ref{Selberg}) is the contribution of the identity element,
${\rm Vol}(\Gamma\backslash G)$ is the (finite) volume of
the fundamental domain with respect to the measure $d\mu$,
$\psi(s)$ is the logarithmic derivative of the Euler $\Gamma-$function,
and $C$ is a computable real constant
\cite{Elizalde,Bytsenko0,Bytsenko3}.
The function $S(s)$ is given by a generalised Dirichlet series,
convergent for $\Re\,s>1$, $
S(s)=\pi^{1/2}\Gamma(s-1/2)[\Gamma(s)]^{-1} \sum_{c\neq
0}\sum_{0\leq d<|c|} |c|^{-2s} \mbox{,} $ where the sums are taken
over all pairs $c,d$ of the matrix
$\left(\begin{array}{ll}*\,\,*\\
c\,\,d\end{array}\right)\subset\Gamma_\infty
\backslash\Gamma/\Gamma_\infty$. Also the poles of the meromorphic
function $S(s)$ are contained in the region $\mathop{\Re}\nolimits
s<1/2$ and in the interval $[1/2,1]$.

\section{Concluding remarks}

{\bf Finite volume cosets $\Gamma\backslash G/K$ and Killing
spinors}. In the previous sections we have discussed supergravity
solutions involving anti de Sitter and real hyperbolic space
factors. Hyperbolic spaces have infinite volume with respect to
the Poincar\'e metric. Thus there are no normalizable modes for
any field configurations in hyperbolic spaces. On the other hands
non-empty bulk and boundary field theories can be obtained by
forming the coset spaces with topology
$\Gamma\backslash {\mathbb H}^N$.

The hyperbolic manifolds ${\mathbb H}^N$, as factors in solution
(i) of Sec. 2, admit Killing spinors. However, having
the space forms (ii), (iii) of Sec. 2: $AdS_{3}\!\times\!{\mathbb
H}^{2}\!\times\!{\mathbb H}^2\!\times\!{\mathbb S}^4$,
$AdS_{2}\!\times\!{\mathbb H}^{2}\!\times\!{\mathbb H}^3\!\times\!
{\mathbb S}^4$, as the solutions of supergravity theory,
one can recognize
that the factors ${\mathbb H}^{2}\!\times\!{\mathbb H}^2$,
${\mathbb H}^{2}\!\times\!{\mathbb H}^3$ cannot leave any unbroken
supersymmetry. Indeed, the following result holds.

\begin{proposition} (T. Friedrich \cite{Friedrich})\,\,\,
A Riemannian spin
manifold $(M^N, g)$ admitting a Killing spinor $\psi \neq 0$ with
Killing number $\mu \neq 0$ is locally irreducible.
\end{proposition}
\noindent {\bf Proof}. Let the locally Riemannian product has the
form $M^N = M^K\times M^{N-K}$. Let ${\mathcal X}, {\mathcal Y}$
are vectors tangent to $M^K$ and $M^{N-K}$ respectively, and,
therefore, the curvature tensor of the Riemannian manifold $(M^N,
g)$ is trivial. Since $\psi$ is a Killing spinor the following
equations hold (see also Eq. (\ref{sf})):
\begin{eqnarray}
&&\nabla_{\mathcal X}\psi = \mu {\mathcal X} \cdot \psi,
\,\,\,\,\,\,\,
4\mu^2 =  [N(N-1)]^{-1}R
\nonumber \\
&& {\rm at}\,\,\,{\rm each}
\,\,\,{\rm point}\,\,\,{\rm of}\,\,\, {\rm a}\,\,\, {\rm
connected}\,\,\, {\rm Riemannian}
\,\,\, {\rm spin} \,\,\, {\rm manifold} \,\,\, (M^N, g)
\label{Kil}
\mbox{,}
\end{eqnarray}
where $R$ is a scalar curvature. Because of (\ref{Kil}) we have
\begin{eqnarray}
&& \nabla_{{\mathcal X}} \nabla_{\mathcal Y}\psi =
\mu(\nabla_{\mathcal X} {\mathcal Y}) \cdot\psi+\mu^2{\mathcal
Y}\cdot{\mathcal X}\cdot \psi \Longrightarrow
\nonumber \\
&& (\nabla_{\mathcal X}\nabla_{\mathcal Y} - \nabla_{\mathcal
Y}\nabla_{\mathcal X} - \nabla_{[{\mathcal X}, {\mathcal Y}]})\psi
= \mu^2({\mathcal Y}\cdot{\mathcal X} -{\mathcal X}\cdot{\mathcal
Y})\psi \label{nabla} \mbox{.}
\end{eqnarray}
The curvature tensor $R({\mathcal X},{\mathcal Y})$ in the spinor
bundle ${\mathfrak S}$ is related to the curvature tensor of the
Riemannian manifold \\
$(M^N, g)$:
$ R({\mathcal X},{\mathcal Y}) =
(1/4)\sum_{j=1}^{N}e_j R({\mathcal X},{\mathcal Y})e_j\cdot \psi,
$ where $\{e_j\}_{j=1}^N$ is a orthogonal basis in manifold.
Therefore Eq. (\ref{nabla}) can also be written as
\begin{equation}
\sum_{j=1}^{N}e_j R({\mathcal X},{\mathcal Y})e_j\psi
+[N(N-1)]^{-1}R ({\mathcal X}{\mathcal Y} - {\mathcal Y}{\mathcal
X})\psi = 0 \label{curv} \mbox{.}
\end{equation}
From Eq. (\ref{curv}) we get $R\cdot {\mathcal X}\cdot {\mathcal
Y}\cdot \psi =0$, and moreover ${\mathcal X}$ and ${\mathcal Y}$
are orthogonal vectors. Since $\mu \neq 0$ ($R \neq 0$) it follows
that $\psi =0$, hence a contradiction. ${\Box}$
\\

Simple type IIB supergravity backgrounds with D3 brane charge and
constant dilaton are the following: $AdS_3\times {\mathbb
H}^2\times {\mathbb S}^5$, $AdS_2\times {\mathbb H}^3\times
{\mathbb S}^5$. Like $AdS_5\times {\mathbb S}^2\times {\mathbb
S}^3$, these spaces have no Killing spinors and solution is not
supersymmetric \cite{Russo}. But solution is regular everywhere
(being a direct product of Einstein spaces), and $\alpha'$
corrections can be made very small for sufficiently large radius
of compact part. Although non-supersymmetric, these spaces is an
interesting setup for string compactification, in particular, for
the construction of conformal field theories. Other
compactifications can be obtained by replacing hyperbolic space
factors ${\mathbb H}^4$ by any Einstein space $X^4$ of the same
negative curvature; it gives the solution $AdS_3 \times X^4 \times
{\mathbb S}^4$. In fact we have the following statement:

\begin{proposition} (T. Friedrich \cite{Friedrich})\,\,\,
Let $(M^N,g)$ be a connected Riemannian spin manifold and let
$\psi$ is a non-trivial Killing spinor with Killing number $\mu
\neq 0$. Then $(M^N,g)$ is an Einstein space.
\end{proposition}
{\bf Rroof}. The proof easily follows from Proposition 2; indeed
$(M^N,g)$ is an Einstein space of scalar curvature given by Eq.
(\ref{Kil}). $\Box$
\\

A question of interest is whether supersymmetry survives under the
orbifolding by the discrete group $\Gamma$. That question could be
addressed also to the factors $\Gamma \backslash ({\mathbb H}^{2}
\!\times\!{\mathbb H}^2)$, $\Gamma \backslash ({\mathbb H}^{2}
\!\times\!{\mathbb H}^3)$. Perhaps there are more complicate
solutions involving real hyperbolic spaces, where some
supersymmetries are unbroken. However analysis of that problem is
complicate and we leave it for other occasion. In fact,
supersymmetry guarantees the stability of the physical system. But
its absence does not necessarily implies instability.
In general, a definite statement for
the stability of supergravity solutions needs a study of the
spectral properties of operators, which has not been well studied
yet.
\\
\\
{\bf String-supergravity correspondence}. A version of duality
states that string theory (or M theory) compactified on spaces of
the form $AdS_{D+1}\times (\Gamma \backslash{\mathbb H}^N) \times
{\mathbb S}^{K}$ defines a $D-$dimensional conformal field theory
with $SO(K+1)$ global symmetry. In addition, the correlation
functions of the conformal field theory are defined as follows
\cite{gkp,WW} $ \langle \exp \{\int d^Dx \ \Phi_{\rm boundary}(x)
{\mathcal O}(x)\} \rangle _{\rm CFT}\equiv {Z}_{\rm string}(\Phi_{\rm
boundary}(x)) \ , $ where ${Z}_{\rm string}$ is the partition
function of the string theory computed with boundary values of the
string fields, which act as sources of operators of conformal
theory. In the classical limit of supergravity the string
partition function can be evaluated as follows: ${Z}_{\rm
string}\cong \exp [-I_{SG}(\Phi)]$, where solution of the
equations of motion in the background $AdS_{D+1}\times
(\Gamma\backslash {\mathbb H}^N) \times {\mathbb S}^{K}$ with the
boundary condition $\Phi_{\rm boundary}$, has to be taken into
account.

Following lines of \cite{Russo} we mention here two interesting
cases of supergravity soilutions. First of them is the $D=11$
supergravity solution of the form $AdS_5\times M^2\times {\mathbb
S}^4$. This model is dual to a $D=4$ non-supersymmetric conformal
field theory with $SO(5)$ global symmetry group. The supergravity
solution has M5-brane charge and one may expect the theory to be
related to the six-dimensional $(2,0)$ or $(1,0)$ conformal field
theories. The second case is the eleven-dimensional supergravity
solution $AdS_4\times M^3\times {\mathbb S}^4$, and the type IIB
solution $AdS_3\times M^2\times {\mathbb S}^5$. The $AdS_4\times
M^3\times {\mathbb S}^4$ solution has M5 brane charge. Thus, the
dual field theory is a $D=2+1$ conformal field theory associated
with the $6D$ $(2,0)$ conformal theory, with an internal global
symmetry group $SO(5)$. Finally the later solution has D3 brane
charge and it should be dual to some ${\cal N}=0$ $D=1+1$
conformal field theory related to ${\cal N}=4$ $D=3+1$ super
Yang-Mills theory.
\\
\\
{\bf Results on the holographic principle}. The backgrounds
considered in previous sections can be used for the construction
of new ${\cal N}=0$ conformal field theories by holography.
According to the holographic principle, there exist strong ties
between certain field theories on a manifold (``bulk space'') and
on its boundary (at infinity). A few mathematically exact results
relevant to that program are the following. The class of Euclidean
AdS$_3$ spaces which we have considered here are quotients of the
real hyperbolic space ${\mathbb H}^3$ by a Schottky group. The
boundaries of these spaces can be compact oriented surfaces with
conformal structure (compact complex algebraic curves).

In \cite{Manin}, a principle associated with the Euclidean AdS$_2$
holography has been established. The bulk space is there a modular
curve, which is the global quotient of the hyperbolic plane
${\mathbb H}^2$ by a finite index group, $\Gamma$, of
$G = PSL(2, {\mathbb Z})$. The boundary at infinity is then
${P}^1({\mathbb R})$. Let $M$
be a coset space $M = \Gamma\backslash G$. Then, the modular curve
$X_{\Gamma}:= \Gamma\backslash {\mathbb H}^2$ can be presented as the
quotient $X_{\Gamma} = G\backslash({\mathbb H}^2\times M)$; its
non-commutative boundary (in the sense of Connes \cite{Connes}) as
the $C^*-$ algebra $C({P}^1({\mathbb R})\times M){>\!\!\!\lhd} G$,
Morita equivalent to $C({P}^1({\mathbb  R})){>\!\!\!\lhd}\Gamma$
\cite{Mar,Manin}. The results which have been regarded as
manifestations of the holography principle are \cite{Manin}:

\vskip .25cm

\begin{itemize}
\item{} There is a correspondence between the eigenfunctions of
the transfer operator $L_s$ and the eigenfunctions of the
Laplacian (Maas wave forms).
\item{} The cohomology classes in $H_1(X_{\Gamma},
{\rm cusps}, {\mathbb  R})$ can be regarded as elements in the cyclic
cohomology of the algebra $C({P}^1({\mathbb  R})\times M){>\!\!\!\lhd}
G$. Cohomology classes of certain geodesics in the bulk space
correspond to projectors in the algebra of observables on the
boundary space.
\item{} An explicit correspondence exists between
a certain class of fields in the bulk space (Mellin transforms of
modular forms of weight two) and the class of fields on the
boundary.
\end{itemize}
Other constructions associated with the symmetric space can be
considered for convex cocompact groups. In fact, let $\partial X$
be a geodesic boundary of the symmetric space $X$ of a real, rank
one, semisimple Lie group $G$. If $\Gamma \subset G$ is a discrete
torsion-free subgroup, then a $\Gamma-$equivalent decomposition,
$\partial X =\Omega \cup \Lambda$, can be constructed, where
$\Lambda$ is the limit set of $\Gamma$. The subgroup $\Gamma$ is
called convex cocompact if $\Gamma\backslash X\cup \Omega$ is a
compact manifold with boundary \cite{Patterson}. The geometry boundary
of ${\mathbb H}^N$ in half-space Poincar\'{e} model is
$\partial_{\infty}{\mathbb H}^N = {\mathbb R}^{N-1}\cup{\infty}$.
If $\Gamma$ is convex cocompact and torsion free, then the orbit
space, $X_{\Gamma}=\Gamma\backslash {\mathbb  H}^{N}$, may be viewed
as the interior of a compact manifold with boundary, namely the
Klein manifold for $\Gamma$, ${\overline X} =
(\Gamma\backslash {\mathbb  H}^{N})\cup (\Gamma\backslash
\Omega(\Gamma))$, so that the boundary at infinity is given by
$\partial_{\infty}X_{\Gamma} =
\partial {\overline{X}} = \Gamma\backslash \Omega(\Gamma)$.
\vspace{5mm}

\subsection*{Acknowledgements}

The authors would like to thank the Conselho Nacional de
Desenvolvimento Cient\'{\i}fico e Tecnol\'ogico (CNPq) for a
support.

\end{document}